\documentclass{article}

\usepackage[english]{babel}

\usepackage[letterpaper,top=2cm,bottom=2cm,left=3cm,right=3cm,marginparwidth=1.75cm]{geometry}

\usepackage{amsmath}
\usepackage{graphicx}
\usepackage[colorlinks=true, allcolors=blue]{hyperref}
\usepackage{cite} 
\usepackage[numbers,sort&compress]{natbib}
\title{Engineering Cryogenic FETs: Addressing SCEs and Impact of Interface Traps Down to 2 K Temperature \\
(Version - 2)}
\author{Nilesh Pandey, \textit{Senior Member, IEEE}, Dipanjan Basu, \\ and Sanjay K. Banerjee, \textit{ Life Fellow IEEE} }

\begin{document}
\maketitle

\begin{abstract}
This paper presents the design and benchmarking of cryogenic bulk-FETs using an experimentally calibrated TCAD framework that integrates 2-D electrostatics and interface-trap effects from \textit{T} = 2~K to 300~K. For a 28-nm node device, carrier transport is predominantly ballistic at \textit{T} = 2~K and becomes quasi-ballistic as temperature increases.
At cryogenic temperatures, higher interface-trap densities increase the effective threshold voltage and suppress subthreshold conduction. However, when the ON-state bias is adjusted to account for the trap-induced $V_t$ shift, interface traps are found to \emph{worsen} $I_{\mathrm{ON}}/I_{\mathrm{OFF}}$ along with degrading the subthreshold swing (SS) and reducing mobility across all temperatures.
The spatial standard deviation $\sigma$ of the trap distribution modulates these behaviors: highly localized traps ($\sigma \sim 1$–$2$~nm) exacerbate short-channel effects (SCEs), whereas broader, nearly uniform distributions ($\sigma \ge 50$~nm) elevate the entire barrier and suppress SCEs until saturation as $\sigma \to L_g$. The TCAD predictions closely match experimental data at 4.2~K, 77~K, and 300~K, providing design guidelines to optimize $I_{\mathrm{ON}}/I_{\mathrm{OFF}}$, SS, mobility, and DIBL for cryogenic CMOS technology nodes.\end{abstract}

\section{Introduction}
Conventional semiconductor FETs are inherently limited by a theoretical limit of subthreshold swing of approximately 60 mV/dec at 300 K, dictated by the high energy tail of the Boltzmann carrier distribution, which constrains computing efficiency at room temperature \cite{streetman2000solid}.
Operating the conventional semiconductor FETs at cryogenic temperatures offers a promising avenue to address the state-of-the-art computational need by potentially surpassing the fundamental switching limit of 60 mV/dec \cite{Beckers}. Additionally, silicon qubit devices and circuits have been explored for quantum computing, further necessitating the reliable operation of devices at cryogenic temperatures \cite{Johns},\cite{Philips2022}.

Research on cryogenic MOSFETs can be broadly divided into three domains: 1) Numerical Simulations (TCAD), 2) experimental characterization, and 3) Modeling. A few TCAD-based works include: Ab Initio atomic simulations \cite{PhysRevApplied.18.054089}, Si mobility model TCAD at cryogenic temperature \cite{Dhillon},  Si - Quantum dot simulations \cite{Pillarisetty},\cite{PandeyDQD} \cite{Mohiyaddin} \cite{pandey2} . Examples of characterizations at cryogenic temperatures include \cite{Beckers} \cite{Ekanayake} \cite{Incandela} \cite{Bonen}. On the modeling perspective: A surface potential-based model is reported in \cite{Incandela}, 1-D Poisson’s equation analytical models in \cite{Beckers_1}, \cite{MARTIN2011115}, and Verilog-A-based models in \cite{Akturk}, \cite{Kabao}, \cite{Zhang}.

  \begin{figure}[!t]
		\centering \hspace{-1mm}
		\includegraphics[width=0.6\textwidth]{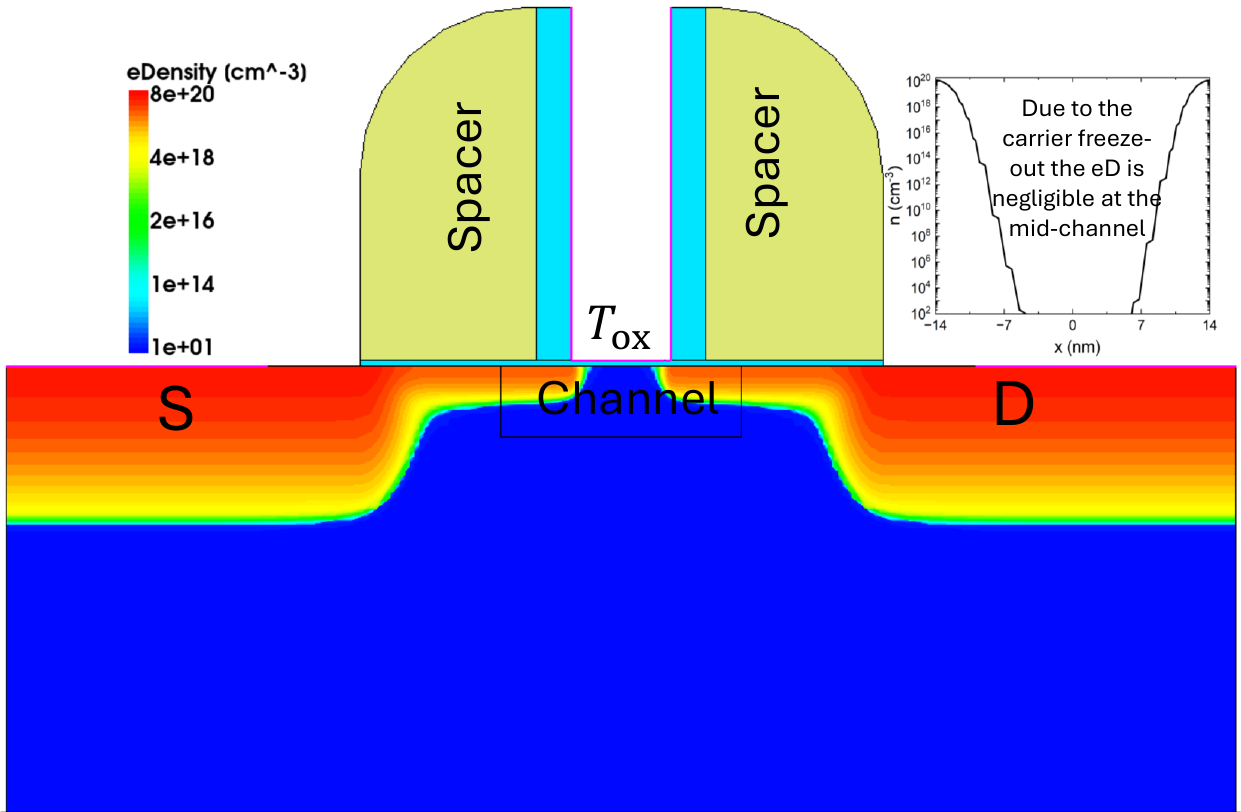}
		\caption{Schematic bulk MOSFET at cryogenic temperature (\( T = 2 \, \text{K} \)). Due to carrier freeze-out, the electron density is negligible at the mid-channel (\( x = 0 \)). Default parameters in this work: Drain bias (\( V_{\text{ds}} \)) = 0 V, Oxide thickness (\( T_{\text{ox}} \)) = 1.5 nm, Gate length (\( L_{\text{g}} \)) = 28 nm, S/D doping = \(6 \times 10^{20} \, \mathrm{cm^{-3}}\), S/D extension region doping = \(2 \times 10^{20} \, \mathrm{cm^{-3}}\), Net S/D doping =  \(8 \times 10^{20} \, \mathrm{cm^{-3}}\) gate-metal work function (\( \phi_m \)) = 4.5 eV, substrate thickness = 1 \(\mu\)m, and Device width = 300 nm.
}\label{fig:schm} 
	\end{figure}
   
This work focuses on the engineering of silicon-based cryogenic FETs through a comprehensive TCAD modeling framework that incorporates 2-D electric field effects, short-channel effects, and integration of cryogenic physics. The study examines the impact of technology node scaling, specifically gate length and oxide thickness, while accounting for interfacial localized defects and uniform trap distributions. 

At cryogenic temperatures, device behavior is strongly influenced by interface traps in addition to short-channel electrostatics. A higher trap density introduces negative charge at the $\text{SiO}_2/\text{Si}$ interface, raising the conduction-band barrier and shifting the threshold voltage upward. Although this suppresses the OFF-state current and can make $I_{\text{ON}}/I_{\text{OFF}}$ appear larger under a fixed-bias comparison, this effect is only apparent and disappears once the ON-state gate voltage is adjusted to account for the trap-induced $V_t$ shift. In reality, the increased trap capacitance adds in parallel with the depletion capacitance, weakening gate control and worsening the subthreshold swing (SS), while also reducing mobility. Consequently, interface traps ultimately degrade device performance at cryogenic temperatures despite any apparent fixed-bias enhancement in $I_{\text{ON}}/I_{\text{OFF}}$.

The spatial distribution of traps also plays a critical role. Highly localized defects, represented by narrow distributions with small standard deviation (\(\sigma\)), perturb the barrier only locally and therefore exacerbate short-channel effects (SCEs), leading to poorer SS and stronger \(V_t\) roll-off. In contrast, a broader trap distribution (\(\sigma \gg L_g\)) acts more uniformly across the interface, raising the entire barrier and partially suppressing SCEs until the effect saturates as \(\sigma\) approaches the channel length.

A 28 nm CMOS device with gate oxide thickness $\sim$ 1.5 nm does not show significant gate tunneling at cryogenic temperatures, as confirmed by experimental studies 
\cite{Beckers}, \cite{Beckers_1}. Therefore, gate tunneling is not considered in this work. However, for more aggressively scaled oxide thicknesses, one may need to include tunneling effects by modeling wavefunction penetration \cite{Banerjeee2}, and quantumization effects \cite{Banerjeee1}, which would require advanced quantum transport simulations, which are beyond the scope of this work.

\section{TCAD Modeling Framework: A Short-Channel Model at Cryogenic Temperature}
\begin{figure*}[!t]
		\centering 
		\includegraphics[width=1\textwidth]{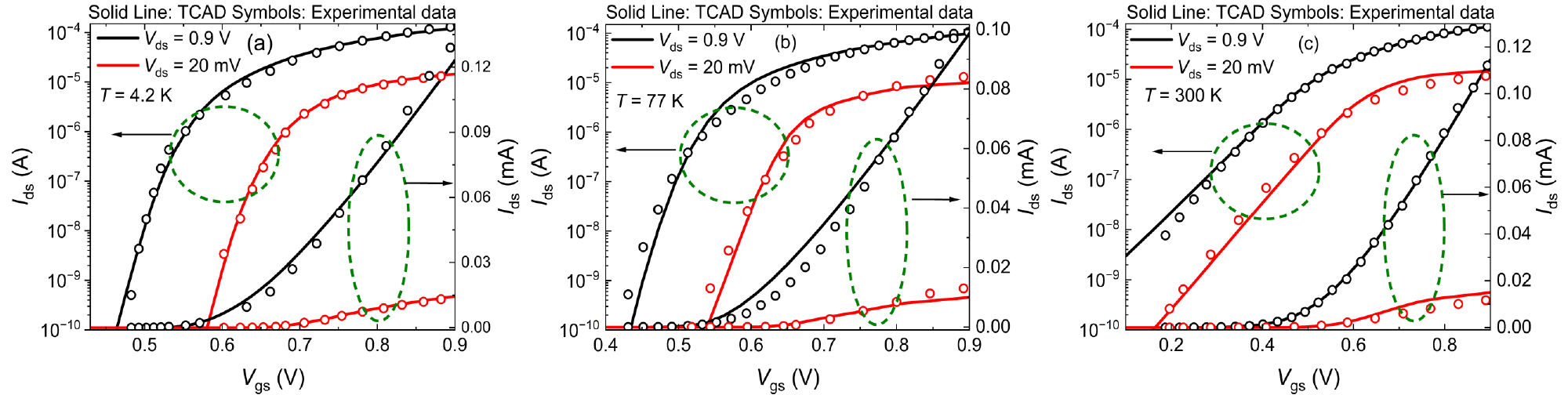}
		\caption{TCAD deck calibration with the experimental data reported in \cite{Beckers}. The ballistic model best fits across the cryogenic temperatures, with \( L_{\text{ch}} \) as a ballistic prefactor defined in the TCAD module. At \( T = 4.2 \, \text{K} \), nearly all carriers exhibit ballistic transport (\( L_{\text{ch}} = 12 \, \text{nm} \)), transitioning to quasi-ballistic transport at \( T = 77 \, \text{K} \) (\( L_{\text{ch}} = 42 \, \text{nm} \)), and negligible ballistic transport at \( T = 300 \, \text{K} \) (\( L_{\text{ch}} \sim 200 \, \text{nm} \)). Note that the \( L_{\text{ch}} \) is not the gate length but a ballisticity parameter defined in the TCAD. Device width = 300 nm.
		}\label{fig:valid}  
	\end{figure*}
Fig. \ref{fig:schm} shows a 28 nm node bulk n-MOS, and Table \ref{tab:device_params} shows the key parameters used in this work. Note that in this work, we do not perform characterization of the FET. Instead, the TCAD deck is calibrated against experimental data available in the literature.
To model cryogenic FET via TCAD Sentaurus, we focus on the three key points \cite{TCAD}.

\begin{table}[h!]
\centering
\begin{tabular}{|c|c|c|}
\hline
Parameter Name & Value   \\ \hline
Gate-Oxide thickness   &   1.5 nm (no gate stack)\\ \hline
Device Width   &   300 nm \\ \hline
S/D Length  &   100 nm \\ \hline
 Spacer Length & 60 nm   \\ \hline
 Poly gate thickness & 100 nm   \\ \hline

  Substrate thickness & 1 $\mu$ m   \\ \hline
 Substrate doping (constant)  & 1.5 $\times$ 10$^{18}$cm$^{-3}$   \\ \hline 
 S/D doping profile & Gaussian, peak value  6 $\times$ 10$^{20}$cm$^{-3}$  \\ \hline
 S/D extensions  & Gaussian, peak value  2 $\times$ 10$^{20}$cm$^{-3}$   \\ \hline
 Net S/D extensions doping  & Gaussian, peak value  8 $\times$ 10$^{20}$cm$^{-3}$   \\ \hline
 Interface trap concentration  & Mentioned in the figures   \\ \hline
\end{tabular}
\caption{Key structural and doping parameters used in the TCAD simulations.}
\label{tab:device_params}
\end{table}

\subsection{Physics Models}
The TCAD simulations incorporate the \texttt{BandTailDOS}, \texttt{QuantumPotential}  model along with quasi-ballistic mobility effects to capture non-equilibrium carrier transport. Interface traps are modeled using a spatially varying Gaussian distribution, which allows for a transition between localized and extended defect states, depending on the standard deviation. The defect profile was chosen to be consistent with experimental observations \cite{streetman2000solid},\cite{Chang},\cite{Bennett}.

\subsection{Numerical Implementation}
Ensuring stable convergence of the Newton solver required careful tuning of key numerical parameters. The default value for the quantum potential was set sufficiently high to maintain self-consistency, while additional stabilization terms were introduced as needed, such as \texttt{ExtendedPrecision} and \texttt{error parameter (ErrRef)} parameters. The solver configuration leverages precision controls and dynamic step adjustments to handle steep field variations and ensure smooth convergence across a wide temperature range.

\subsection{Solution Strategy}
The simulation follows a multi-step transient process:  
\begin{enumerate}
\item \textbf{Initialization:} Gate and drain bias are ramped gradually using transient ramping at $T = 300$ K to establish a stable operating point.  
\item \textbf{Temperature Modeling:} A second transient simulation introduces temperature-dependent electrostatic and transport effects by dynamically adjusting bias-dependent parameters. This allows for a seamless transition to cryogenic conditions without artificial discontinuities.  
\item \textbf{Final Sweep:} A controlled voltage sweep is performed to capture the complete electrostatic response of the device, ensuring consistency with prior biasing conditions.  
\end{enumerate}

Fig. \ref{fig:valid} validates the TCAD deck against the experimental data reported in Ref.~\cite{Beckers} over temperatures from 4.2~K to 300~K.

\section{Cryogenic Electrostatics}

\begin{figure*}[!t]
		\centering 
		\includegraphics[width=1\textwidth]{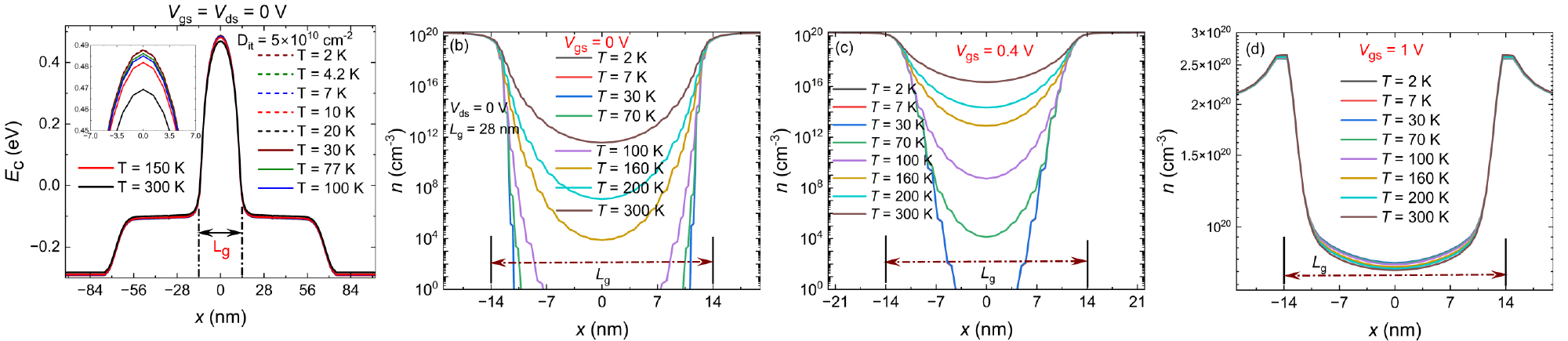}
		\caption{ (a) Conduction band barrier plotted at the interface for the various temperatures. (b) At zero gate bias, the carrier density increases significantly from \( T = 2 \, \text{K} \) to \( T = 300 \, \text{K} \), driven by the change in occupation probability and shows relatively negligible dependency on the change in surface potential. At very low temperatures (\( T < 30 \, \text{K} \)), the carrier density follows \( n \propto \exp \left( -\frac{1}{2k_B T} \right) \). 
(c) \& (d) For \( V_{\text{gs}} > 0.4 \, \text{V} \), saturation in charge density occurs at higher temperatures, while at low temperatures, the saturation boundary is less defined.
		}\label{fig:ed_Vg} 
	\end{figure*}

Fig. \ref{fig:ed_Vg} shows the electron density near the Si/SiO\(_2\) interface. At zero drain bias, the channel's Fermi level aligns with the source/drain Fermi levels, as shown in Fig. \ref{fig:ed_Vg}(a). The maximum surface potential changes only by 25 mV with temperature [Fig. \ref{fig:ed_Vg}(a)], but the carrier density increases by nearly 10\(^ {11}\) in order of magnitude from \( T = 2 \, \text{K} \) to \( T = 300 \, \text{K} \) [Fig. \ref{fig:ed_Vg}(b)]. At very low temperatures (\( T < 30 \, \text{K} \)) with the Fermi level well below the channel band edge, the carrier density in an undoped semiconductor at zero drain bias is governed by \( n \propto \exp\left(-\frac{1}{2k_B T}\right) \) \cite{Tyagi}. This indicates that the primary driver of the carrier population with temperature is the thermal energy (\( k_B T \) term) rather than changes in surface potential (at zero biases). It is important to note that using \( n = n_i \exp\left(\frac{q \phi}{k_B T}\right) \) to explain carrier density would lead to incorrect conclusions, as it relies on the Maxwell-Boltzmann approximation, which is invalid at cryogenic temperatures requiring the full Fermi-Dirac distribution. As the gate voltage (\( V_{\text{gs}} \)) increases, more carriers populate in the channel due to reduced barrier height. At temperatures between \( T = 200 \, \text{K} \) and 300 K, saturation in charge density is observed for \( V_{\text{gs}} \geq 0.4 \, \text{V} \), see Fig. \ref{fig:ed_Vg}(b)-(d). However, the saturation range lacks a well-defined boundary at very low temperatures.
\begin{figure}[!t]
		\centering 
		\includegraphics[width=1\textwidth]{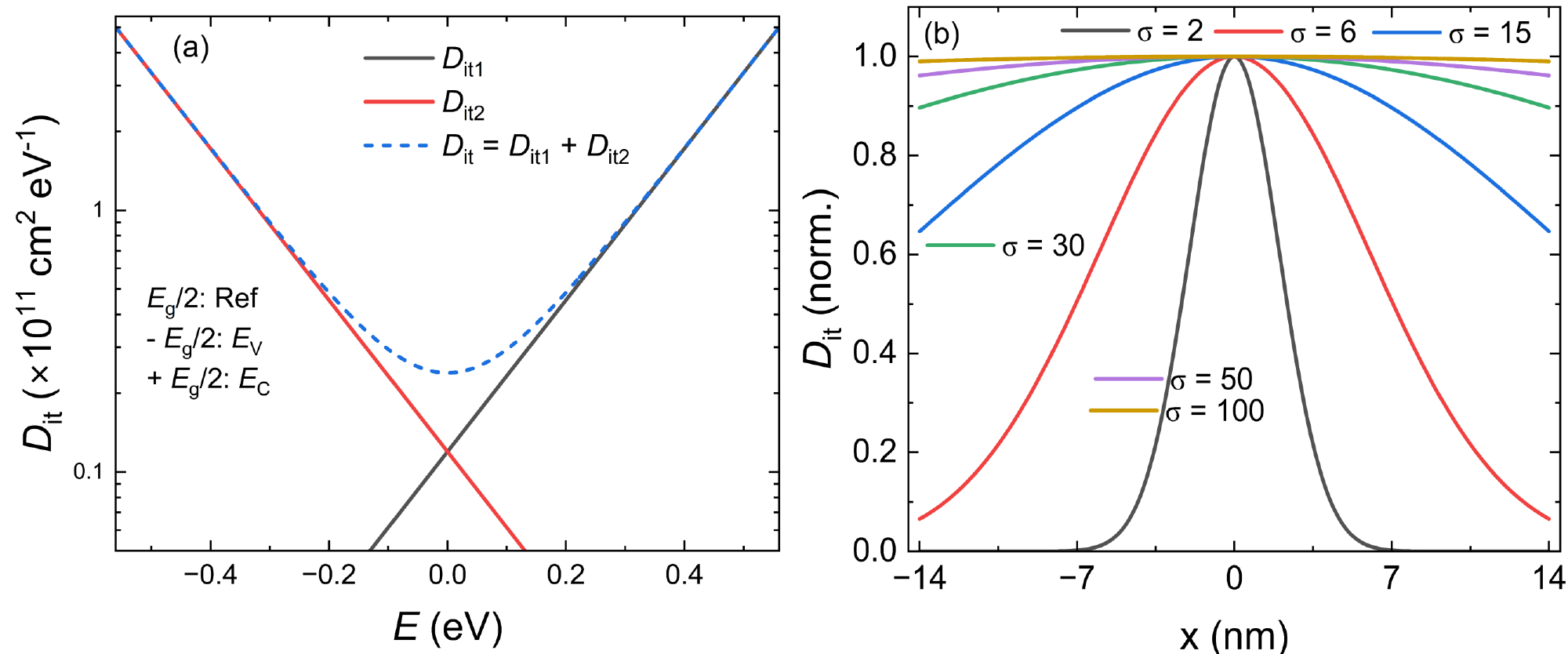}
		\caption{(a) Trap density-of-states $D_{\mathrm{it}}(E)$ considered in the TCAD simulations, plotted over the entire Si bandgap. The U-shaped distribution extends from the valence-band edge ($E_V = -E_g/2$) to the conduction-band edge ($E_C = +E_g/2$), with the energy axis referenced to midgap. (b) Spatial interface trap distribution is defined as a Gaussian distribution with varying standard deviation to model a highly localized trap profile ($\sigma$ = 2 nm and uniform trap distribution $\sigma \geq$ 50 nm.
		}\label{fig:trap} 
	\end{figure}

\section{Trap Density and Cryogenic Transport: Length Scaling and SCEs}
The interface trap density $D_{\text{it}}$ at the Si/SiO$_2$ interface is modeled with a realistic U-shaped energy dependence, consistent with experimental observations, where $D_{\text{it}}$ peaks near the conduction ($E_c$) and valence ($E_v$) band edges and reaches a minimum near mid-gap \cite{sze2007physics},\cite{White}. Fig. \ref{fig:trap}(a) shows the energy dependence of $D_{\text{it}}(E)$ used in the TCAD simulations. Two exponential trap distributions are used to effectively capture the U-shaped distribution. 
The spatial distribution of traps along the channel is modeled as a Gaussian function with varying standard deviation $\sigma$, as shown in Fig. \ref{fig:trap}(b), to capture both highly localized ($\sigma = 1~\mathrm{nm}$) and nearly uniform trap profiles ($\sigma \geq 50~\mathrm{nm}$). 

The spatial shape of the traps is defined by utilizing TCAD's key words \texttt{SpatialShape}, \texttt{SpaceSig}, and \texttt{SpaceMid} \cite{TCAD}. These trap states are occupied when an electron is captured with a negative charge and unoccupied when uncharged.

The minimum trap density is \( D_{\text{it}} = 5 \times 10^{10} \, \text{cm}^{-2} \), and the maximum trap density is \( D_{\text{it}} = 5 \times 10^{11} \, \text{cm}^{-2} \), which falls within the range of state-of-the-art silicon-based cryogenic FETs \cite{Pillarisetty}, \cite{Paul}. A higher trap density corresponds to a non-ideal interface, which can be observed in experiments \cite{streetman2000solid}. Therefore, we consider this range of trap densities to account for all possible scenarios, from ideal to worst-case conditions.

\begin{figure*}[!t]
		\centering 
		\includegraphics[width=1\textwidth]{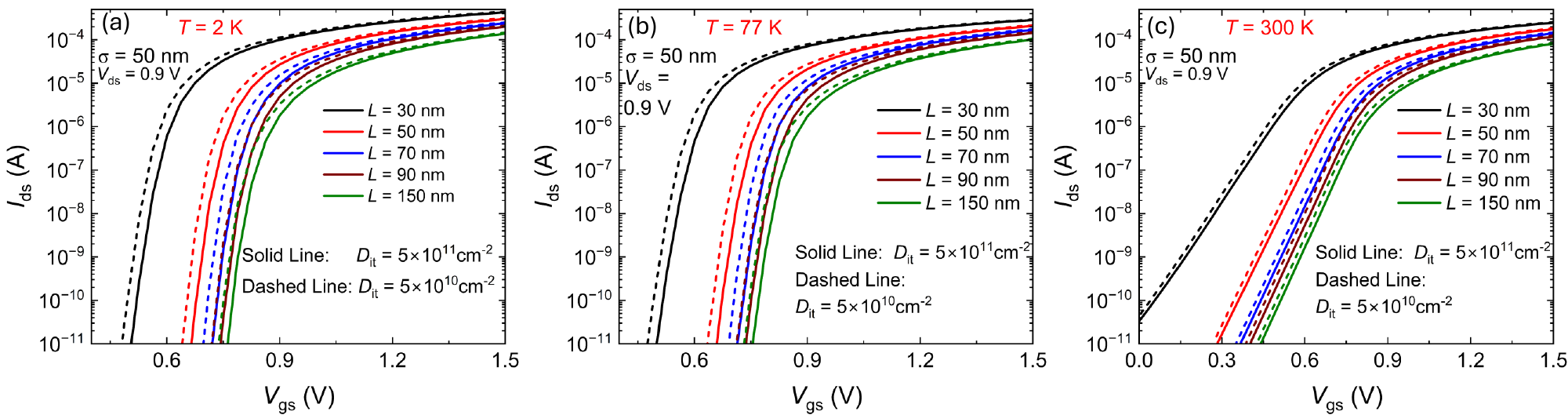}
		\caption{
(a)–(c) Transfer characteristics of devices with different channel lengths ($L_{\text{g}} = 30$–$150~\text{nm}$) at $T = 2$, 77, and 300~K, using a Gaussian interface-trap distribution with $\sigma = 50~\text{nm}$ ($>\!L_{\text{g}}$). Solid and dashed curves correspond to $D_{\text{it}} = 5\times10^{11}\,\text{cm}^{-2}$ and $5\times10^{10}\,\text{cm}^{-2}$, respectively. A higher trap density raises the threshold voltage due to additional negative charge at the $\text{SiO}_2/\text{Si}$ interface. Under a fixed-bias comparison, this $V_t$ shift can produce an \emph{apparent} increase in the $I_{\text{ON}}/I_{\text{OFF}}$ ratio at cryogenic temperatures. However, this effect disappears once the ON-state gate voltage is corrected for the trap-induced $V_t$ shift, and interface traps are found to degrade $I_{\text{ON}}/I_{\text{OFF}}$. Device width = 300~nm.
}
\label{fig:Id_Vg_L_Dit_Temp} 
	\end{figure*}

Fig.~\ref{fig:Id_Vg_L_Dit_Temp} uses a Gaussian trap profile with $\sigma = 50~\text{nm}\,(>L_g)$, which makes $D_{\mathrm{it}}$ nearly uniform along the $\mathrm{SiO_2/Si}$ interface [see Fig.~\ref{fig:trap}(b)]. Increasing $D_{\mathrm{it}}$ introduces additional negative charge at the interface, raising the electrostatic barrier and shifting the threshold voltage $V_t$ upward. To quantify this effect, we first extract $V_t$ for both high- and low-$D_{\mathrm{it}}$ devices by locating the gate voltage at which $I_{\mathrm{OFF}} = 10^{-9}~\text{A}$. The resulting threshold difference $\Delta V_t = V_{t,\mathrm{high}} - V_{t,\mathrm{low}}$ is then used to correct the ON-state bias for the high-$D_{\mathrm{it}}$ case, i.e., $I_{\mathrm{ON,low}} = I_d(V_g = 1.0~\text{V})$ and $I_{\mathrm{ON,high}} = I_d(V_g = 1.0~\text{V} - \Delta V_t)$. This bias correction removes any artificial boost in $I_{\mathrm{ON}}/I_{\mathrm{OFF}}$ that would arise solely from the upward $V_t$ shift.

After correcting the ON-state bias for the high-$D_{\mathrm{it}}$ device by subtracting $\Delta V_t$, its $I_{\mathrm{ON}}/I_{\mathrm{OFF}}$ ratio becomes lower than the low-$D_{\mathrm{it}}$ case by factors of approximately 0.835 (2~K), 0.857 (77~K), and 0.884 (300~K). Thus, the apparent enhancement observed under fixed-bias comparison originates entirely from the trap-induced $V_t$ shift and not from a true performance improvement.

         \begin{figure*}[!t]
		\centering 
		\includegraphics[width=1\textwidth]{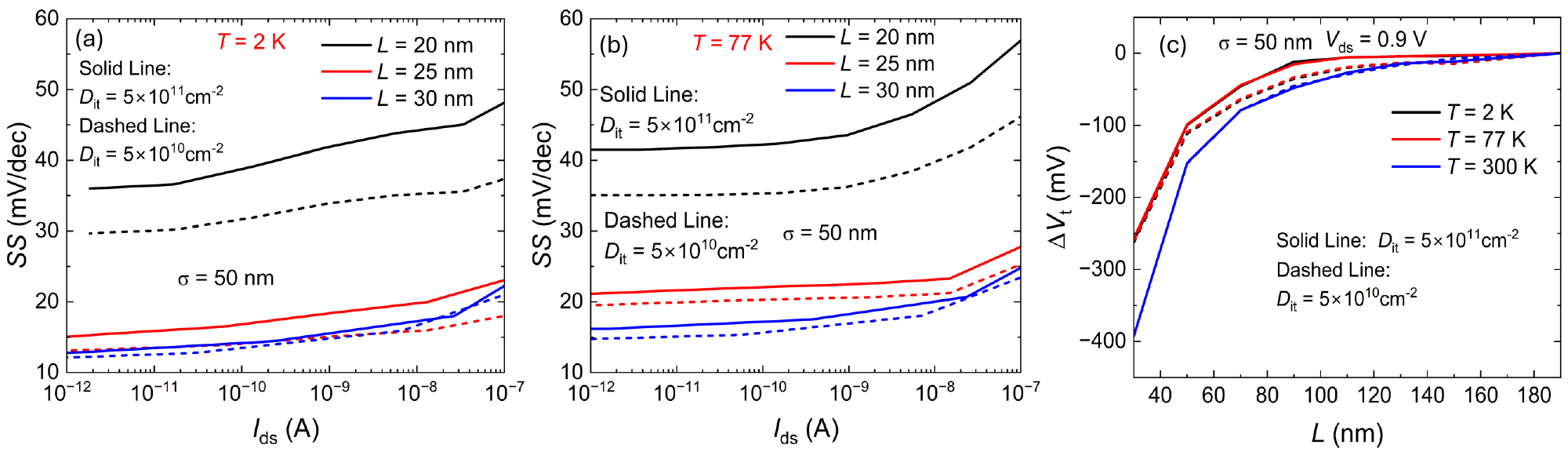}
		\caption{(a) \& (b) Subthreshold swing (SS) as a function of drain current for \(L_{\text{g}}=20{-}30~\text{nm}\) at \(T=2~\text{K}\) and \(T=77~\text{K}\) comparing devices with high (\(D_{\text{it}}=5\times10^{11}\,\text{cm}^{-2}\), solid) and low (\(D_{\text{it}}=5\times10^{10}\,\text{cm}^{-2}\), dashed) trap densities. Higher trap density increases SS, degrading the switching steepness. The difference is most evident at \(L=20~\text{nm}\) due to stronger SCEs, but it reduces with increasing \(L\). (c) Threshold voltage shift \(\Delta V_t\) versus channel length for \(T=2,\,77,\) and \(300~\text{K}\). Increasing \(L_{\text{g}}\) raises \(V_t\) and saturates beyond \(L_{\text{g}}\approx150~\text{nm}\), consistent with the long-channel limit. The roll-off trends for high and low \(D_{\text{it}}\) are nearly identical, since trap-induced \(V_t\) shifts are comparable at both short and long channel lengths.
.}\label{fig:SCE_Dit_sigma_high} 
	\end{figure*}
\begin{figure*}[!t]
		\centering 
		\includegraphics[width=1\textwidth]{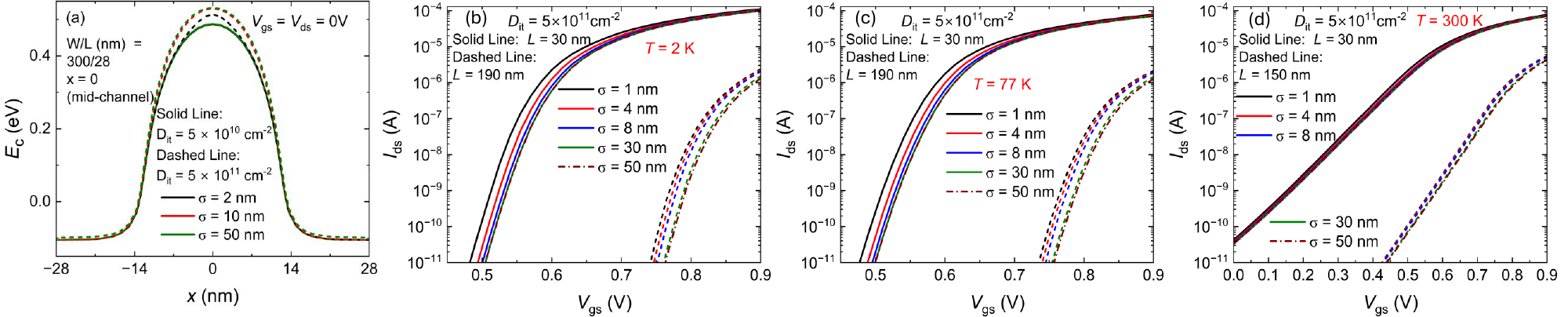}
		\caption{(a) Conduction band profile at the channel interface showing the effect of trap distribution width \((\sigma=2{-}50~\text{nm})\) for low \(\left(D_{\text{it}}=5\times10^{10}~\text{cm}^{-2}\right)\) and high \(\left(D_{\text{it}}=5\times10^{11}~\text{cm}^{-2}\right)\) trap densities. A broader distribution increases the effective barrier height until saturation as \(\sigma\) approaches the gate length. (b)–(d) Transfer characteristics at \(T=2,\,77,\) and \(300~\text{K}\) for \(L=30~\text{nm}\) (solid) and long channel limit (dashed), showing the dependence on \(\sigma\). At cryogenic temperatures, the impact of \(\sigma\) is pronounced: uniform distributions (\(\sigma \gg L\)) raise the entire conduction band and suppress short-channel effects, whereas localized traps (\(\sigma \approx 2~\text{nm}\)) only perturb the barrier locally, resulting in degraded OFF-state control. Device width = 300 nm.
		}\label{fig:Id_Vg_Sigma_Dit_L_28nm_Temp} 
	\end{figure*}

Fig.~\ref{fig:SCE_Dit_sigma_high} illustrates the influence of interface trap density on the subthreshold swing (SS) and threshold voltage roll-off. A higher trap density \(\left(D_{\text{it}}=5\times10^{11}\,\text{cm}^{-2}\right)\)  increases SS due to capacitance via trap charges, thereby degrading the switching speed following the literature \cite{streetman2000solid}. 
This degradation is more pronounced at \(T=77~\text{K}\) compared to \(T=2~\text{K}\), since additional thermal carriers at elevated temperature increase the subthreshold swing, as shown in Fig.~\ref{fig:SCE_Dit_sigma_high} (a) and (b). 
Furthermore, for very short channel lengths (\(L_{\text{g}}=20~\text{nm}\)), the difference in SS between high and low \(D_{\text{it}}\) becomes clearly visible, highlighting the strong impact of short-channel effects (SCEs) in amplifying the role of SS caused by trap states [see Fig. \ref{fig:SCE_Dit_sigma_high}(a)].
\begin{figure*}[!t]
		\centering 
		\includegraphics[width=0.85\textwidth]{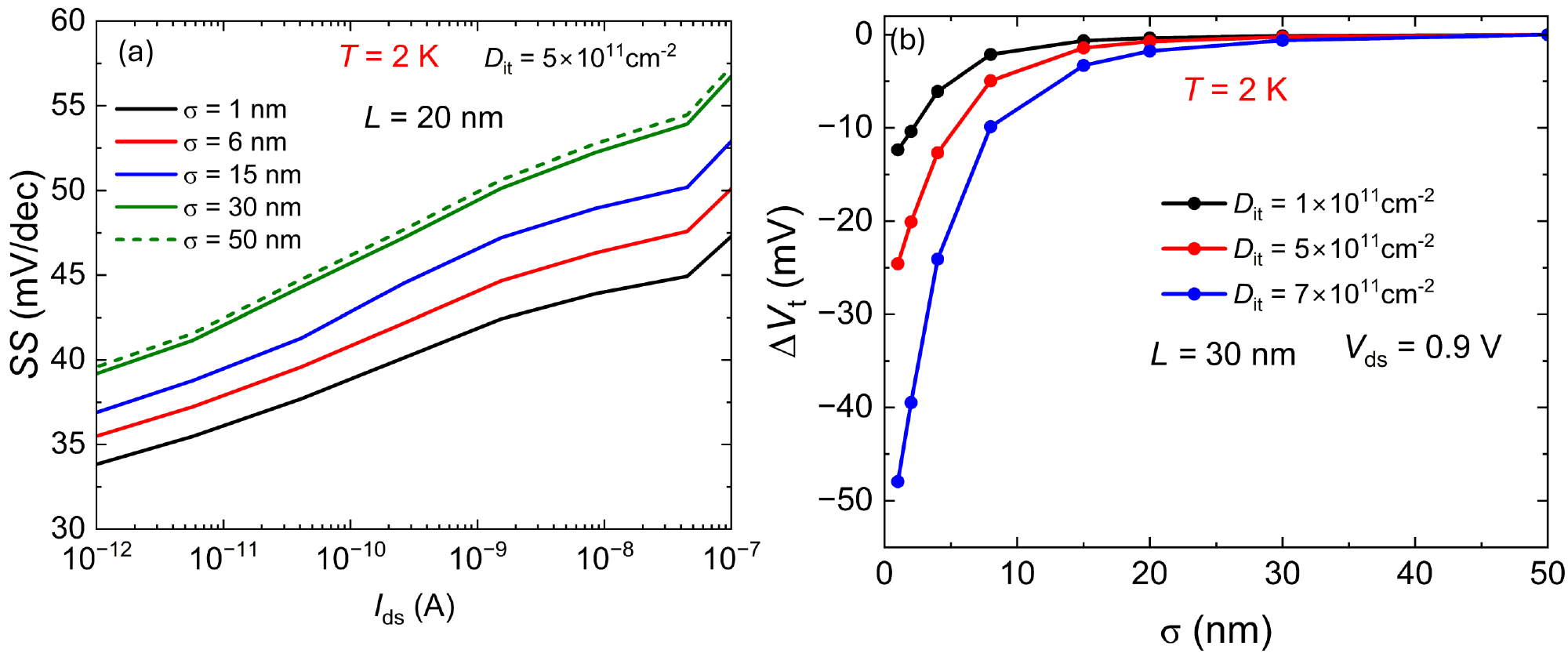}
		\caption{ (a) Subthreshold swing (SS) as a function of drain current for \(L=20~\text{nm}\) at \(T=2~\text{K}\), showing the impact of increasing trap distribution standard deviation for \(D_{\text{it}}=5\times10^{11}~\text{cm}^{-2}\). Larger \(\sigma\) values correspond to broader effective trap areas, which degrade SS due to the additional trap-state capacitance. As \(\sigma\) approaches the channel length, SS saturates since the total trap density no longer increases with \(\sigma\). (b) Threshold voltage shift versus \(\sigma\), showing similar saturation behavior. Stronger roll-off is observed for small \(\sigma\), while for large \(\sigma\) the shift saturates. Higher trap density \((7\times10^{11}~\text{cm}^{-2})\) demonstrates more severe degradation in both SS and \(V_t\).
		}\label{fig:SCE_Dit_Temp} 
	\end{figure*}
The threshold-voltage roll-off characteristics in Fig.~\ref{fig:SCE_Dit_sigma_high}(c) show that increasing $L_{g}$ produces an upward shift in $V_t$. For $L_{g} \ge 150~\text{nm}$, this shift saturates, indicating that $L_{g} \approx 150~\text{nm}$ represents the long-channel limit. The roll-off for high and low $D_{\mathrm{it}}$ are nearly identical because a higher trap density introduces a similar absolute $V_t$ shift for both short ($L_{g}=30~\text{nm}$) and long ($L_{g}=150~\text{nm}$) channels, thereby preserving the relative roll-off trend.
A similar behavior is observed at $T = 2~\text{K}$ and $T = 77~\text{K}$, where the roll-off curves nearly overlap. At these temperatures, the trap-induced $V_t$ shift is dominant, with the same magnitude, and short-channel effects are strongly suppressed, resulting in comparable roll-off characteristics. In contrast, at $T = 300~\text{K}$ the roll-off becomes significantly stronger due to increased short-channel effects arising from the larger population of thermally activated carriers, which enhances barrier lowering and degrades electrostatic control.

 The subthreshold swing (SS) is extracted at each gate voltage and plotted as a function of the drain current. In this representation, devices compared at the same current level correspond to the same surface potential, which enables a fair comparison of their switching characteristics. The threshold voltage roll-off is determined using the constant-current method at a reference current of \(10~\text{nA}/\mu\text{m}\).


    \section{Impact of Localized Defect (trap) on the Transport \& SCEs}

 Fig.~\ref{fig:Id_Vg_Sigma_Dit_L_28nm_Temp} shows the impact of the standard deviation \((\sigma)\) of the trap distribution on the conduction band profile and transfer characteristics at different temperatures. A narrow distribution (\(\sigma = 2~\text{nm}\)) represents a highly localized defect, while a broad distribution (\(\sigma = 50~\text{nm}\)) corresponds to a nearly uniform trap density across the entire channel interface. Increasing \(\sigma\) broadens the trap-induced charge distribution, which enhances the overall barrier modulation. As \(\sigma\) approaches the gate length, the barrier height saturates because the total trap charge in the channel also saturates, as shown in Fig. \ref{fig:Id_Vg_Sigma_Dit_L_28nm_Temp}(a). 
 
 At higher trap density \(\left(D_{\text{it}}=5\times10^{11}~\text{cm}^{-2}\right)\), the influence of the trap distribution is especially visible in the OFF-state at cryogenic temperatures (\(T=2\)\;–\;77~K), where a pronounced deviation is observed between localized defects (\(\sigma = 1~\text{nm}\)) and uniform distributions (\(\sigma = 50~\text{nm}\)), as shown in Fig. \ref{fig:Id_Vg_Sigma_Dit_L_28nm_Temp}(b) - (d). A uniform distribution raises the entire conduction band profile, leading to improved suppression of short-channel effects, while a localized trap only perturbs the barrier locally, resulting in weaker electrostatic control. 
 

Fig. \ref{fig:SCE_Dit_Temp} shows the effect of the trap distribution width (\(\sigma\)) on the subthreshold swing (SS). The increasing \(\sigma\) effectively broadens the trap distribution area at the interface, which increases the effective trap density experienced by the channel. This leads to a degradation of SS, since the additional trap-state capacitance appears in parallel with the depletion capacitance, thereby reducing the overall gate control. As \(\sigma\) approaches the channel length, a saturation behavior is observed because the trap density becomes uniformly distributed across the interface and does not increase further with \(\sigma\). A similar saturation trend is also seen in the threshold voltage roll-off, where lower values of \(\sigma\) exhibit stronger roll-off due to localized trapping, but converge as \(\sigma\) increases, as shown in Fig. \ref{fig:SCE_Dit_Temp}(b). To assess more extreme conditions, a higher trap density of \(7\times10^{11}~\text{cm}^{-2}\) is included, which shows even stronger degradation in threshold voltage roll-off. 

    \section{Impact of Oxide Thickness Scaling and DIBL}
Fig. \ref{fig:Ec_DIBL_Tox_T_4} shows the impact of drain bias on the conduction band barrier height at \( T = 4.2 \, \text{K} \) for various oxide thicknesses. The reduction in the maximum barrier height under high drain bias is approximately \( 70 \, \text{mV} \) and \( 95 \, \text{mV} \) for \( T_{\text{ox}} = 1.5 \, \text{nm} \) and \( 2 \, \text{nm} \), respectively. A higher oxide thickness increases the drain-induced barrier lowering (DIBL) effect due to reduced gate control.

\begin{figure*}[!t]
    \centering
    \includegraphics[width=1\textwidth]{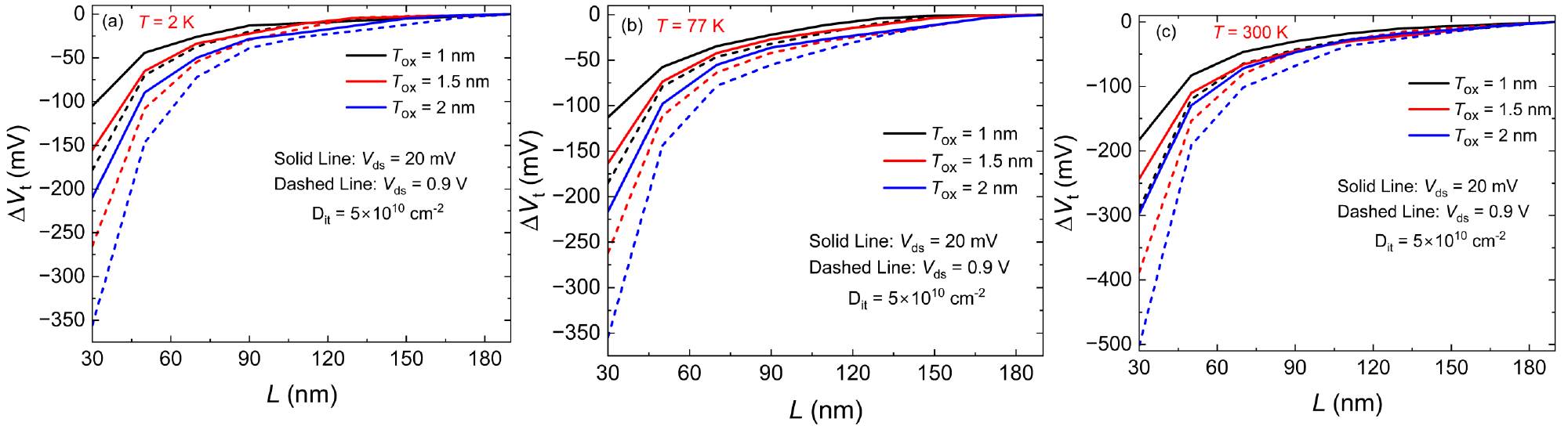}
    \caption{(a)-(c) Threshold voltage roll-off and DIBL as a function of channel length for various oxide thicknesses. Increasing oxide thickness reduces gate control, resulting in a more pronounced DIBL effect. The impact of DIBL is quantified by \(\Delta V_t|_{V_{ds} = 20 \, \text{mV}} - \Delta V_t|_{V_{ds} = 0.9 \, \text{V}}\), which is approximately \( 72 \, \text{mV} \), \( 98 \, \text{mV} \), and \( 142 \, \text{mV} \) for \( T_{\text{ox}} = 1 \, \text{nm} \), \( 1.5 \, \text{nm} \), and \( 2 \, \text{nm} \), respectively, at \( L_g = 30 \, \text{nm} \) and \( T = 2 \, \text{K} \). The effect increases significantly at \( T = 300 \, \text{K} \), as shown in Fig. (c).}
    \label{fig:DIBL_Temp}
\end{figure*}

Fig. \ref{fig:DIBL_Temp} shows the threshold voltage roll-off and DIBL as a function of channel length for various oxide thicknesses. Increasing oxide thickness reduces gate control over the device, resulting in a more pronounced DIBL effect, as expected. The impact of DIBL is quantified by analyzing \(\Delta V_t|_{V_{ds} = 20 \, \text{mV}} - \Delta V_t|_{V_{ds} = 0.9 \, \text{V}}\), which is approximately \( 72 \, \text{mV} \), \( 98 \, \text{mV} \), and \( 142 \, \text{mV} \) for \( T_{\text{ox}} = 1 \, \text{nm} \), \( 1.5 \, \text{nm} \), and \( 2 \, \text{nm} \), respectively, at \( L_g = 30 \, \text{nm} \) and \( T = 2 \, \text{K} \). Similar values are observed at \( T = 77 \, \text{K} \). However, with the temperature change from \( T = 77 \, \text{K} \) to \( T = 300 \, \text{K} \), the magnitude of threshold voltage roll-off significantly increases for each oxide thickness. Furthermore, the DIBL at \( T = 300 \, \text{K} \) is approximately \( 107 \, \text{mV} \), \( 145 \, \text{mV} \), and \( 205 \, \text{mV} \) for \( T_{\text{ox}} = 1 \, \text{nm} \), \( 1.5 \, \text{nm} \), and \( 2 \, \text{nm} \), respectively, at \( L_g = 30 \, \text{nm} \).

It is also noted that the impact of interface trap density on DIBL is negligible in this case, as a relatively small value of \(D_{\text{it}} = 5\times10^{10}~\text{cm}^{-2}\text{eV}^{-1}\) is considered, corresponding to an interface that closely approximates the ideal limit.
\begin{figure}[!t]
		\centering \hspace{-3mm}
		\includegraphics[width=0.4\textwidth]{Figures/Ec_DIBL_Tox_T_4.2K.png}
		\caption{Impact of drain bias on conduction band barrier height at \( T = 4.2 \, \text{K} \) for the various oxide thicknesses. The maximum barrier height reduction under high drain bias is \( 70 \, \text{mV} \) and \( 95 \, \text{mV} \) for \( T_{\text{ox}} = 1.5 \, \text{nm} \) and \( 2 \, \text{nm} \), respectively, with a higher oxide thickness enhancing the DIBL effect due to reduced gate control.
		}\label{fig:Ec_DIBL_Tox_T_4}
	\end{figure}
\begin{figure*}[!t]
	\centering
	\includegraphics[width=1\textwidth]{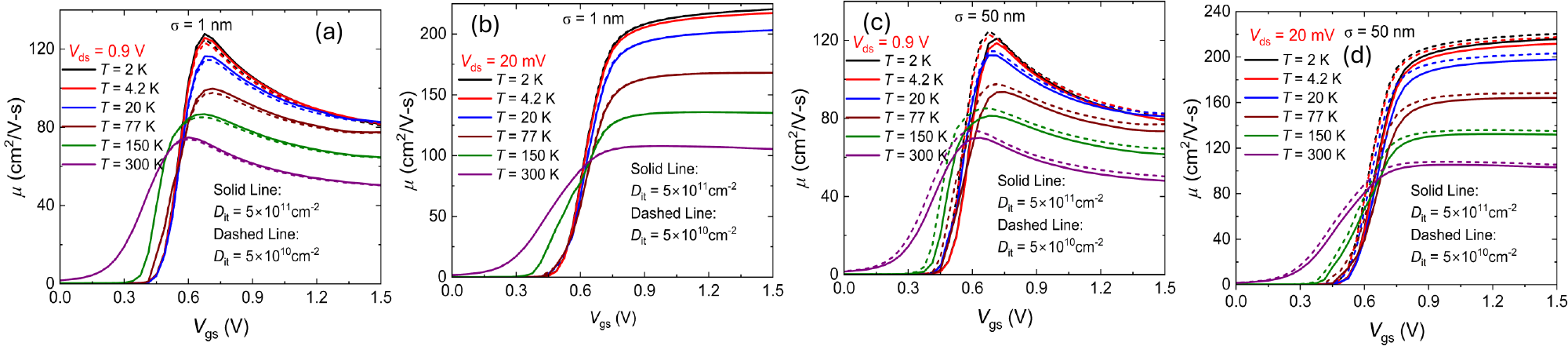}
	\caption{(a)–(d) Extracted effective surface mobility versus gate voltage for different temperatures (\(T=2{-}300~\text{K}\)) and trap distribution widths (\(\sigma=1~\text{nm}\) and \(50~\text{nm}\)), comparing low drain bias (\(V_{\text{ds}}=20~\text{mV}\)) and high drain bias (\(V_{\text{ds}}=0.9~\text{V}\)). Solid and dashed lines represent high \((D_{\text{it}}=5\times10^{11}\,\text{cm}^{-2})\) and low \((D_{\text{it}}=5\times10^{10}\,\text{cm}^{-2})\) trap densities, respectively. At small drain bias, mobility increases and saturates with \(V_{\text{g}}\) due to carrier-induced screening of interface traps. At high drain bias, significant degradation from peak to high-\(V_{\text{g}}\) mobility is observed, particularly at cryogenic temperatures. Broader trap distributions (\(\sigma=50~\text{nm}\)) reduce the peak mobility compared to localized traps (\(\sigma=1~\text{nm}\)), as the scattering centers are distributed uniformly across the interface.
	}\label{fig:Mob_sigma_Dit_Temp_Vds}
\end{figure*}
   \section{Cryogenic Mobility and Trap Distribution}
Fig.~\ref{fig:Mob_sigma_Dit_Temp_Vds} shows the impact of trap distribution width (\(\sigma\)) and drain bias on the effective electron surface mobility from cryogenic to room temperature. At a small drain bias of \(V_{\text{ds}}=20~\text{mV}\) [Figs.~\ref{fig:Mob_sigma_Dit_Temp_Vds}(b) and (d)], the mobility increases monotonically with gate voltage and saturates at higher bias. At low gate voltages, the carrier density is very small, resulting in poor screening of charged impurities and interface traps. Coulomb scattering therefore dominates and severely limits mobility. As the gate voltage increases, the accumulated carriers screen the scattering centers, leading to a rapid improvement in mobility and eventual saturation. The saturation mobility clearly decreases with temperature, dropping from above \(220~\text{cm}^2/\text{V·s}\) at \(T=2~\text{K}\) to below \(110~\text{cm}^2/\text{V·s}\) at \(T=300~\text{K}\), indicating a nearly twofold reduction across this range.  

The effect of the trap distribution width becomes evident when comparing localized traps (\(\sigma=1~\text{nm}\)) with nearly uniform distributions (\(\sigma=50~\text{nm}\)) [Figs.~\ref{fig:Mob_sigma_Dit_Temp_Vds}(a) and (c)]. A larger \(\sigma\) corresponds to a broader effective trap area at the interface, which increases surface scattering across the entire channel and reduces the peak mobility. For example, at \(T=2~\text{K}\) and high drain bias (\(V_{\text{ds}}=0.9~\text{V}\)), the maximum mobility decreases from about \(124~\text{cm}^2/\text{V·s}\) for \(\sigma=1~\text{nm}\) to \(118~\text{cm}^2/\text{V·s}\) for \(\sigma=50~\text{nm}\). Furthermore, under large drain bias, mobility degradation at high gate voltages is more pronounced, especially at cryogenic temperatures, where the percentage drop from peak to high-\(V_{\text{g}}\) mobility is significantly larger.


Note that the plotted mobility represents the average effective mobility, computed using the Effective Mobility physical model interface (PMI) in Sentaurus Device, integrated along the channel cross-section. Although the absolute values may differ from experimental extractions due to methodological differences, the simulated bias and temperature trends remain consistent with measurements.

\section*{Conclusion}
A comprehensive TCAD study of bulk MOSFETs from 2 to 300~K reveals that interface traps and short-channel electrostatics jointly determine cryogenic device behavior. At low temperatures, trap-induced barrier raising suppresses subthreshold conduction and can create an \emph{apparent} enhancement in the $I_{\mathrm{ON}}/I_{\mathrm{OFF}}$ ratio under fixed-bias comparison. However, once the ON-state gate voltage is adjusted to account for the trap-induced threshold shift, this apparent improvement vanishes, and interface traps are found to degrade $I_{\mathrm{ON}}/I_{\mathrm{OFF}}$, subthreshold swing, and mobility across all temperatures.
Trap density and its spatial distribution further modulate device behavior. Localized defects increase the barrier non-uniformly and worsen $I_{\mathrm{OFF}}$, whereas broader or nearly uniform trap distributions (larger $\sigma$) raise the entire barrier and mitigate SCEs at cryogenic temperatures. Channel lengths beyond 150~nm yield saturation in $V_t$ and SS, identifying this as the long-channel limit for cryogenic bulk-FET operation.
These findings underscore the importance of optimizing interface-trap profiles and geometric scaling to preserve electrostatic integrity and transport at low temperatures. For high-performance cryogenic CMOS, minimizing interface-trap density is especially critical, and device design must incorporate the effects of carrier freeze-out in non-degenerately doped channel regions.

\newpage
\section*{Appendix: Mobility Modeling in TCAD}

The carrier mobility in our simulations is modeled using a combination of physically rigorous models implemented in \texttt{SDevice}. The complete syntax used is:

\begin{quote}
\texttt{Mobility (}\\
\quad\texttt{DopingDependence ( PhuMob BalMob(Lch = @Lch\_0@) )}\\
\quad\texttt{Enormal (IALMob)}\\
\quad\texttt{HighFieldSaturation}\\
\texttt{)}
\end{quote}

\subsection*{Philips Unified Mobility Model (PhuMob)}

The Philips Unified Mobility model (\texttt{PhuMob}) captures two major contributions to mobility using Matthiessen’s rule:
\begin{equation}
    \frac{1}{\mu_{i,b}} = \frac{1}{\mu_{i,L}} + \frac{1}{\mu_{i,\mathrm{DAeh}}}
\end{equation}
Here, \( \mu_{i,L} \) accounts for phonon scattering:
\begin{equation}
    \mu_{i,L} = \mu_{i,\mathrm{max}} \left(\frac{T}{300\,\mathrm{K}}\right)^{-\zeta_i}
\end{equation}
and \( \mu_{i,\mathrm{DAeh}} \) includes scattering from ionized dopants \cite{TCAD}, free carriers, and clustering effects. The screening parameter \( P_i \), which determines \( G(P_i) \) and \( F(P_i) \), depends on total carrier concentration \( (n+p) \), doping, and effective mass, thus indirectly introducing gate-bias dependence into the model \cite{TCAD}.

\subsection*{Ballistic Mobility Model (BalMob)}

At nanoscales, carriers experience quasi-ballistic transport. This is modeled using the injection velocity–dependent ballistic model:
\begin{equation}
    \mu_{\text{bal}} = f(V_{\text{inj}}, T) \cdot L_{\text{ch}}
\end{equation}
where $L_{ch}$ is the ballistic parameter, not the channel length, which is tuned to calibrate the experimental data. The function \( f \) captures the injection velocity, drain/source bias, and temperature dependencies:
\begin{equation}
    f = \frac{1}{V_{\mathrm{ds}} + V_b} \cdot \frac{V_b}{V_{\mathrm{ds,lin}}} \left(\frac{L_{\text{ch}}}{L_{\text{ref}}}\right)^{\alpha} \cdot \frac{V_{\text{inj}} \left[1 - \exp\left(-\frac{V_{\mathrm{ds}}}{k_BT}\right)\right]}{1 + \exp\left(-\frac{V_{\mathrm{ds}}}{k_BT}\right)}
\end{equation}
The ballistic mobility is added to the total mobility using Matthiessen's rule and dominates at low temperatures where phonon and impurity scattering are suppressed.

\subsection*{Validation of $L_{\text{ch}}$ for Temperature Mapping}
Net mobility is calculated as
\begin{equation}
    \frac{1}{\mu} = \frac{1}{\mu_{i,b}} + \frac{1}{\mu_{\text{bal}}}
\end{equation}. Therefore, a very large value of $\mu_{\text{bal}}$ (or a very large value of $L_{\text{ch}}$ will have a negligible impact on the net mobility calculation.
The $L_{ch} $values used to model different transport regimes are:
\begin{itemize}
    \item \( T = 4.2\,\mathrm{K} \): \( L_{\text{ch}} = 12\,\mathrm{nm} \) — strongly ballistic,
    \item \( T = 77\,\mathrm{K} \): \( L_{\text{ch}} = 42\,\mathrm{nm} \) — quasi-ballistic,
    \item \( T = 300\,\mathrm{K} \): \( L_{\text{ch}} = 200\,\mathrm{nm} \) — diffusive.
\end{itemize}

These values are physically reasonable because at cryogenic temperatures:
\begin{itemize}
    \item Phonon scattering decreases significantly due to the Bose–Einstein occupation:
    \[
    N_{\text{phonon}} \propto \frac{1}{e^{\hbar \omega / kT} - 1}
    \]
    \item Ionized impurity scattering also weakens due to dopant freeze-out.
\end{itemize}
This leads to enhanced mean free paths and stronger ballistic transport.

\subsection*{Inversion and Accumulation Layer Mobility Model (IALMob)}

The degradation of mobility at the oxide/semiconductor interface due to vertical electric field is modeled using \texttt{IALMob}. This component depends on the normal field \( E_\perp \), which is controlled by gate voltage. Additionally, if a non-uniform trap distribution is defined (e.g., via \texttt{eBandTailDOS(Gaussian)}), then the interface trap density of states (Dit) can also modulate mobility. A broader spatial or energetic distribution (\( \sigma \)) increases Coulomb scattering near the interface, and is accounted for implicitly through \texttt{IALMob} and the carrier distribution near the interface.

Detailed mobility models are explained in the TCAD manual (chapter 15) \cite{TCAD}.
\subsection*{High Field Saturation}

To account for carrier velocity saturation under strong longitudinal fields, we also include:
\begin{quote}
\texttt{HighFieldSaturation}
\end{quote}
This models the reduction in mobility when the longitudinal electric field exceeds a critical threshold.
\subsection*{Impact of Interface Traps on Mobility}

Although the interface trap density (\(D_{\mathrm{it}}\)) does not explicitly appear in the mobility equations, its effect is included implicitly through electrostatics. Trapped charges modify the local surface potential and the vertical electric field (\(E_\perp\)), which in turn alters the inversion charge distribution at the semiconductor/oxide interface. Since field-dependent mobility models such as \texttt{IALMob} and \texttt{PhuMob} use the local electric field and carrier density as primary inputs, any variation caused by traps is automatically reflected in the calculated mobility.
\subsection*{Summary}
Together, this mobility model captures:
\begin{itemize}
    \item Gate-bias effects via field-dependent and charge-screened scattering,
    \item Ballistic to diffusive transition across temperature and channel length,
    \item Interface scattering enhanced by trap distributions,
    \item Velocity saturation at high longitudinal fields.
\end{itemize}

This comprehensive formulation provides a reliable match to experimental data across a wide range of temperatures, channel lengths, and bias conditions.
\section*{Acknowledgment}
The authors gratefully acknowledge Prof. Leonard F. Register and Prof. Yogesh S. Chauhan for their insightful discussions, which strengthened the device physics analysis.


\begin{thebibliography}{99}
\bibitem{streetman2000solid}
B. G. Streetman and S. Banerjee, \textit{Solid State Electronic Devices}, 4th ed., Prentice Hall, New Jersey, 2000, Chapter: Field Effect Transistors.
\bibitem{Beckers}
A. Beckers, F. Jazaeri and C. Enz, "Characterization and Modeling of 28-nm Bulk CMOS Technology Down to 4.2 K," \textit{IEEE Journal of the Electron Devices Society}, vol. 6, pp. 1007–1018, 2018, doi: 10.1109/JEDS.2018.2817458.
\bibitem{Johns}
N. Jone, "Silicon quantum electronics," \textit{Nature Briefing}, vol. 561, no. 3, pp. 163–166, 2018, doi: 10.1038/d41586-018-06610-y.
\bibitem{Philips2022}
S. G. J. Philips et al., "Universal control of a six-qubit quantum processor in silicon," \textit{Nature}, vol. 609, no. 7929, pp. 919–924, 2022, doi: 10.1038/s41586-022-05117-x.
\bibitem{PhysRevApplied.18.054089}
S. Liu et al., "Performance Limit of Gate-All-Around Si Nanowire Field-Effect Transistors: An Ab Initio Quantum Transport Simulation," \textit{Physical Review Applied}, vol. 18, no. 5, p. 054089, 2022, doi: 10.1103/PhysRevApplied.18.054089.
\bibitem{Dhillon}
P. Dhillon and H. Y. Wong, "A Wide Temperature Range Unified Undoped Bulk Silicon Electron and Hole Mobility Model," \textit{IEEE Transactions on Electron Devices}, vol. 69, no. 4, pp. 1979–1983, 2022, doi: 10.1109/TED.2022.3152471.

\bibitem{Pillarisetty}
R. Pillarisetty et al., "Si MOS and Si/SiGe quantum well spin qubit platforms for scalable quantum computing," in \textit{2021 IEEE International Electron Devices Meeting (IEDM)}, 2021, pp. 14.1.1–14.1.4, doi: 10.1109/IEDM19574.2021.9720567.

\bibitem{PandeyDQD}
N. Pandey, D. Basu, L. F. Register and S. K. Banerjee,
"Three-Dimensional Electrostatic and Quantum-Confinement Modeling of Silicon Nanowire Double Quantum Dots,"
arXiv:2510.07831 [cond-mat.mes-hall], 2025,
doi: 10.48550/arXiv.2510.07831.

\bibitem{Mohiyaddin}
F. A. Mohiyaddin, J. P. Slack-Smith, W. Akhtar, R. Rahman, S. Barraud, A. S. Dzurak and A. R. Hamilton,
"Multiphysics Simulation \& Design of Silicon Quantum Dot Qubit Devices,"
in \textit{2019 IEEE International Electron Devices Meeting (IEDM)}, San Francisco, CA, USA, 2019,
pp. 39.5.1--39.5.4,
doi: 10.1109/IEDM19573.2019.8993541.
\bibitem{pandey2}
N. Pandey, D. Basu, Y. S. Chauhan, L. F. Register and S. K. Banerjee, "Engineering Si-Qubit MOSFET: Quantum-Electrostatic Integration at Cryogenic Temperatures," in \textit{IEEE Transactions on Electron Devices}, vol. 72, no. 7, pp. 3881-3888, July 2025, doi: 10.1109/TED.2025.3570285.
\bibitem{Ekanayake}
S. R. Ekanayake et al., "Characterization of SOS-CMOS FETs at Low Temperatures for the Design of Integrated Circuits for Quantum Bit Control and Readout," \textit{IEEE Transactions on Electron Devices}, vol. 57, no. 2, pp. 539–547, 2010, doi: 10.1109/TED.2009.2037381.
\bibitem{Incandela}
R. M. Incandela et al., "Characterization and Compact Modeling of Nanometer CMOS Transistors at Deep-Cryogenic Temperatures," \textit{IEEE Journal of the Electron Devices Society}, vol. 6, pp. 996–1006, 2018, doi: 10.1109/JEDS.2018.2821763.
\bibitem{Bonen}
S. Bonen et al., "Cryogenic Characterization of 22-nm FDSOI CMOS Technology for Quantum Computing ICs," \textit{IEEE Electron Device Letters}, vol. 40, no. 1, pp. 127–130, 2019, doi: 10.1109/LED.2018.2880303.


\bibitem{Beckers_1}
A. Beckers, F. Jazaeri and C. Enz, "Cryogenic MOS Transistor Model," \textit{IEEE Transactions on Electron Devices}, vol. 65, no. 9, pp. 3617–3625, 2018, doi: 10.1109/TED.2018.2854701.
\bibitem{MARTIN2011115}
P. Martin et al., "MOSFET modeling for design of ultra-high performance infrared CMOS imagers working at cryogenic temperatures: Case of an analog/digital 0.18 µm CMOS process," \textit{Solid-State Electronics}, vol. 62, no. 1, pp. 115–122, 2011, doi: 10.1016/j.sse.2011.01.004.
\bibitem{Akturk}
A. Akturk et al., "Compact and Distributed Modeling of Cryogenic Bulk MOSFET Operation," \textit{IEEE Transactions on Electron Devices}, vol. 57, no. 6, pp. 1334–1342, 2010, doi: 10.1109/TED.2010.2046458.
\bibitem{Kabao}
A. Kabaoglu et al., "Statistical MOSFET Modeling Methodology for Cryogenic Conditions," \textit{IEEE Transactions on Electron Devices}, vol. 66, no. 1, pp. 66–72, 2019, doi: 10.1109/TED.2018.2877942.
\bibitem{Zhang}
Y. Zhang et al., "Characterization and Modeling of Native MOSFETs Down to 4.2 K," \textit{IEEE Transactions on Electron Devices}, vol. 68, no. 9, pp. 4267–4273, 2021, doi: 10.1109/TED.2021.3099775.
\bibitem{Banerjeee2}
S. Mudanai et al., "Understanding the effects of wave function penetration on the inversion layer capacitance of NMOSFETs," \textit{IEEE Electron Device Letters}, vol. 22, no. 3, pp. 145–147, 2001, doi: 10.1109/55.910624.

\bibitem{Banerjeee1}
C.-Y. Wu et al., "Quantization effects in inversion layers of PMOSFETs on Si (100) substrates," \textit{IEEE Electron Device Letters}, vol. 17, no. 6, pp. 276–278, 1996, doi: 10.1109/55.496456.
\bibitem{TCAD}
Synopsys Inc., "TCAD Sentaurus, V-2024.03, March 2024," \textit{Synopsys TCAD}, available: https://www.synopsys.com/manufacturing/tcad.
\bibitem{Chang}
W. J. Chang, M. P. Houng and Y. H. Wang, "Simulation of stress-induced leakage current in silicon dioxides: A modified trap-assisted tunneling model considering Gaussian-distributed traps and electron energy loss," \textit{Journal of Applied Physics}, vol. 89, no. 11, pp. 6285–6293, 2001, doi: 10.1063/1.1367399.

\bibitem{Bennett}
H. S. Bennett et al., "Modeling MOS capacitors to extract Si–SiO\textsubscript{2} interface trap densities in the presence of arbitrary doping profiles," \textit{IEEE Transactions on Electron Devices}, vol. 33, no. 6, pp. 759–765, 1986, doi: 10.1109/T-ED.1986.22565.

\bibitem{Tyagi}
M. S. Tyagi, \textit{Introduction to Semiconductor Materials and Devices}, Chapter 3, Wiley, 1991.
\bibitem{sze2007physics}
S. M. Sze and K. K. Ng, \textit{Physics of Semiconductor Devices}, 3rd ed., Wiley, Hoboken, NJ, USA, 2007.
\bibitem{White}
M. H. White and J. R. Cricchi, "Characterization of thin-oxide MNOS memory transistors," \textit{IEEE Transactions on Electron Devices}, vol. 19, no. 12, pp. 1280–1288, 1972, doi: 10.1109/T-ED.1972.17591.



\bibitem{Paul}
A. Paul et al., "Interface trap density metrology from sub-threshold transport in highly scaled undoped Si n-FinFETs," \textit{Journal of Applied Physics}, vol. 110, no. 12, p. 124507, 2011, doi: 10.1063/1.3660697.








\end{thebibliography}
\end{document}